# Electrical sensing of the thermal and light induced spin transition in robust contactless spin-crossover/graphene hybrid devices

*Miguel Gavara-Edo[1], Rosa Córdoba[1]\*, Francisco Javier Valverde-Muñoz[1], Javier Herrero-Martín[2], José Antonio Real[1] and Eugenio Coronado[1]\**

Miguel Gavara-Edo, Dr. Rosa Córdoba, Dr. Francisco Javier Valverde-Muñoz, Prof. José Antonio Real, Prof. Eugenio Coronado
Instituto de Ciencia Molecular, Universidad de Valencia, Catedrático José Beltrán 2, Paterna, 46980, Spain.
E-mail: rosa.cordoba@uv.es, eugenio.coronado@uv.es

Dr. Javier Herrero-Martín
CELLS-ALBA Synchrotron, Carrer de la Llum 2-26, Cerdanyola del Vallès, 08290, Spain

Abstract

Hybrid devices based on spin-crossover (SCO)/2D heterostructures grant a highly sensitive platform to detect the spin transition in the molecular SCO component and tune the properties of the 2D material. However, the fragility of the SCO materials upon thermal treatment, light irradiation or contact with surfaces and the methodologies used for their processing have limited their applicability. Here, we report an easily processable and robust SCO/2D hybrid device with outstanding performance based on the sublimable SCO [Fe(Pyrz)$_2$] molecule deposited over CVD-graphene, which is fully compatible with electronics industry protocols. Thus, a novel methodology based on growing an elusive polymorph of [Fe(Pyrz)$_2$] (tetragonal phase) over graphene is developed that allows us to electrically detect a fast and effective light-induced spin transition in the devices (~50% yield in 5 minutes). Such performance can be enhanced even more when a flexible polymeric layer of PMMA is inserted in between the two active components in a contactless configuration, reaching a ~100 % yield in 5 minutes.

# 1. Introduction

Magnetic switchable molecular materials have recently focused the interest of the scientific community towards their integration in electronic devices.[1,2] Among the different known families constituting this class of molecular compounds, spin crossover (SCO) materials have emerged as promising candidates to this purpose due to their molecular bistability.[3–5] Particularly, the SCO phenomenon has been observed in coordination materials based on Fe(II) ions. In these compounds two different possible electronic configurations may be available, namely the diamagnetic low-spin state (LS, $S = 0$) and the paramagnetic high-spin state (HS, $S = 2$). Interestingly, the transition between these two states can be induced by applying various external stimuli such as temperature, light, pressure, X-ray irradiation, magnetic field and electric field.[6–11] Thus, SCO materials could be utilized as components of diverse functional devices sensitive to these stimuli.

The first attempt to integrate these switching materials into electronic nanodevices was reported by some of us by attaching a single SCO molecular nanoparticle of $[Fe(Htrz)_2(trz)](BF_4)$, where trz = triazole, between two metallic electrodes.[12] A variation in the conductance of the device associated to the thermal spin transition was detected near room temperature. In this case, the conductance ratio between the HS and LS led to an ON/OFF switching ratio of ca. 2.5. Subsequently, we improved this achievement in devices formed by 2D assemblies of these SCO nanoparticles (NPs), where ON/OFF ratios were increased up to two orders of magnitude,[13,14] and finally in assemblies formed by Au/SCO core/shell NPs,[15] where up to three orders of magnitude were obtained.

However, the above devices exhibit low thermal cyclability mainly due to the fragility of the SCO compound when the electrical current is passing through it. An alternative was proposed to overcome this limitation that consists of depositing the SCO nanostructure over graphene, that would be acting as highly sensitive electrical transport component of the hybrid heterostructure.[16] Thus, the SCO NPs were deposited on Chemical Vapor Deposition (CVD) graphene in a four-probe field-effect configuration.[16] In this case, the reversible change in the electrical properties of the graphene upon the thermal spin transition of the NPs probed the device to be a highly-efficient sensor of the spin state that operates near room temperature. In a next step, this electrical sensing was also exploited by covalently attaching [Fe(Htrz)$_2$(trz)](BF$_4$)/SiO$_2$ NPs on chemically exfoliated 2H-MoS$_2$ semiconducting flakes.[17] In this case, the band gap of this semiconductor showed to vary upon the spin transition due to the strain generated in this hybrid heterostructure by the change in volume associated to this phenomenon. This has enabled us to optically detect the spin transition by measuring the photoluminescence of the MoS$_2$.

The previous concept was recently extended to heterostructures combining 2D materials with SCO single crystals.[18,19] Thus, a bulk single-crystal of [Fe(bapbpy)(NCS)$_2$], where bapbpy = N,N′-di(pyrid-2-yl)-2,2′-bipyridine-6,6′-diamine, placed on a CVD graphene layer separated by a polymeric interlayer (poly(methyl methacrylate), PMMA), conferred again thermal bistability to the hybrid device.[18] On the other hand, van der Waals heterostructures formed by mechanically exfoliated thin layers of a SCO crystal ({Fe-(3-Clpyridine)$_2$-[Pt$^{II}$(CN)$_4$]}) and graphene or semiconducting WSe$_2$ layers were prepared.[19] In these, the strain effects induced by the thermal spin transition over the 2D materials were electrically and optically detected.

Notice that all these above-mentioned reported studies were based on the use of SCO compounds, obtained from solution techniques in the form of nanoparticles as well as single-crystals and, furthermore, this spin transition was thermally induced. In order to change this situation, one needs on the one hand to use sublimable SCO molecules, since they are compatible with the processing technologies generally used in electronics,[20] and on the other to use SCO compounds in which the spin transition can be also triggered by light (Light Induced Excited Spin State Trapping, LIESST effect). In this context, very few thermally stable Fe(II) SCO molecules able to maintain their SCO properties upon deposition on surfaces, were reported so far.[21–26] Among them, the LIESST effect was only exploited in two cases: i) vertical devices embedding the SCO molecule [Fe(H$_2$B(pz)$_2$)$_2$(phen)], where H$_2$B(pz)$_2$ = dihydrobis(pyrazolyl)borate and phen = 1,10-phenantroline, in between ITO and Al electrodes and ii) horizontal devices combining graphene with the sublimable SCO molecule [Fe(Pyrz)$_2$], where Pyrz = hydrotris-(3,5-dimethyl-pyrazolyl)borate.[27,28] In the first case, although a fast electrical response to light irradiation from the molecular SCO film was detected, this effect was extremely small.[27] In the second case, the molecular SCO film (130 nm) showed a clear thermal and photoinduced spin transition that was sensed by the change in the electric transport properties of graphene using a two-probe method.[28] However, this photoinduced spin transition was unpractical due to the long irradiation times required (several hours) and the experimental lack of device cyclability studies.

Here we develop a novel protocol to fabricate a series of horizontal switching devices based on hybrid heterostructures formed by CVD graphene and thin layers of sublimable [Fe(Pyrz)$_2$] SCO molecules. The devices exhibit unmatched properties. First, a fast and efficient photoinduced spin transition (quantitative LIESST effect in a few minutes) is induced in the device, which can be electrically detected with high accuracy by using a four-probe configuration. Second, unprecedented robustness of the switching devices upon applying thermal and light stimuli and upon time (more than one month) is demonstrated. Third, the novel procedure has allowed to prepare contactless devices exhibiting an outstandingly enhanced light-induced spin switching, as compared with the thermal one, using PMMA as intermediate layer between graphene and the SCO molecular film.

## 2. Results and Discussion

*Preparation and structural study of the SCO films integrated on the SCO/graphene devices.*
Nanocrystalline continuous molecular films of [Fe(Pyrz)$_2$] with thicknesses ranging from 75 nm to 200 nm were integrated into pre-contacted CVD-graphene horizontal devices (**Suppl. 3. and 4.**). These molecular films were deposited directly on top of the 2D material by sublimation in High Vacuum (HV) of the readily synthesized bulk SCO compound,[7,23,29] forming hybrid electronic devices (**Figure 1a)**. Note that two polymorphs of [Fe(Pyrz)$_2$] have been described, namely the triclinic stable phase and the tetragonal metastable one.[29] However, only the triclinic polymorph has been isolated and studied in detail before. This includes the magnetic characterization and its electrical read-out in hybrid graphene devices.[28–30] The tetragonal polymorph has been an elusive phase which can only be formed by sublimation, appearing always mixed with the triclinic one. In fact, it was proposed that this polymorph is SCO inactive.[29] Here we have developed a novel film growth protocol allowing to directly deposit each polymorph at will.

At this respect, by setting two different temperatures for the substrates (100 °C and -90 °C) during the film deposition, while keeping the same sublimation conditions (molecular source heated at ~145 °C in a ~5·10$^{-8}$ mbar pressure HV and a ~0.4 Å·s$^{-1}$ deposition rate) (**Suppl. 1.2.**), crystalline thin films of the triclinic and the tetragonal polymorphs were separately grown. These films fully retained their chemical integrity according to IR and Raman spectroscopies (**Suppl. 2.1. and 2.2.**) and their crystallographic identity was evidenced by Surface X-Ray Diffraction (SXRD) (**Suppl. 2.4.**). Thus, as-sublimed films deposited at -90 °C revealed the exclusive presence of the tetragonal polymorph (diffraction peaks appearing at 9.95° and 19.97° in 2θ scale; red line in **Figure 1b,c**), while as-sublimed films deposited at 100 °C presented only the ones of the triclinic polymorph (9.92°, 10.04°, 10.15°, 19.97° and 20.42°; blue line in **Figure 1b,c**).[29] Additionally, substantial morphological differences in terms of roughness and crystalline grain sizes were found, being the roughness much lower and the crystals smaller for the tetragonal polymorph films (**Suppl. 2.3.1. and 2.3.2.**). These differences were mainly explained by temperature gradient effects on the nanocrystals growth dynamics, where higher gradients would induce higher nucleation rates and favor precipitation growth regimes, leading to smaller crystals and higher amorphous material fraction constituting the films.[31]

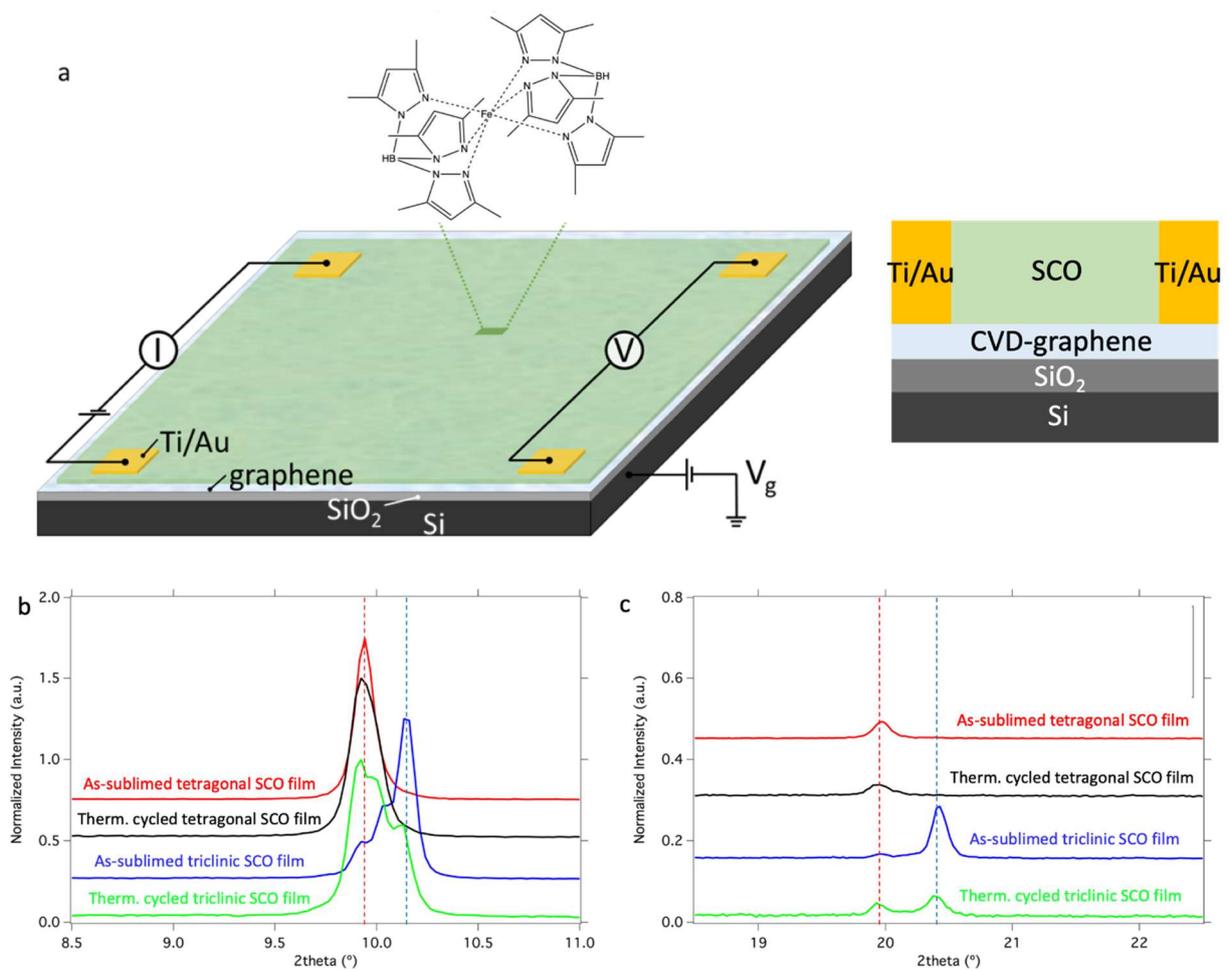

**Figure 1**. a) Scheme of the four-probe DC electrical measurements performed on a fabricated in-contact CVD-graphene horizontal device integrating a [Fe(Pyrz)$_2$] film, in inset the structure of [Fe(Pyrz)] molecule and on the right the profile of the devices. SXRD patterns of [Fe(Pyrz)$_2$] films on CVD graphene-based devices deposited at -90 °C (red line - 250 nm as-sublimed, black line - 120 nm after conductivity measurements) and 100 °C (blue line - 130 nm as-sublimed, green line - 150 nm after conductivity measurements) in 2θ ranges between b) 8.5 ° and 11 ° and c) 18.5 ° and 22.5 °. The colored dashed vertical lines indicate the respective [Fe(Pyrz)$_2$] crystallographic polymorphs most intense diffraction peaks (red – tetragonal and blue – triclinic).

*Synchrotron study of the thermal spin transition in the SCO films.* Before performing the electrical transport measurements of the devices, the SCO behavior of each crystallographic polymorph was analyzed through X-ray Absorption Spectroscopy (XAS) technique at the Fe $L_{2,3}$ edge. For the as-sublimed films of the triclinic polymorph an almost full thermal SCO between 250 K and 90 K is accomplished (~70 % yield; **Figure 2d**), whereas for the as-sublimed films of the tetragonal one a partial spin transition is obtained (~40 % yield; **Figure 2e**). Finally, after performing a few thermal cycles in the range 2-250 K the yield of the SCO undergoes a further decrease (down to ~20 % yield; **Figure 2f**). This study shows that the triclinic polymorph behaves as expected for both the bulk and the post-sublimation annealed films (**Suppl. 2.5.2.**).[9,28–30] As far as the as-sublimed films of the tetragonal polymorph are concerned (**Suppl. 2.5.3.**), the partial thermal SCO transition between 250 K and 90 K is an unexpected result since this polymorph was previously described as SCO inactive.[29] This can be due to the presence of an amorphous SCO component, which has a SCO behavior very close to that of the triclinic polymorph. This component is formed in the film during its growth, as indicated above. Such a phenomenon was extensively studied by measuring XAS for different thicknesses on different surfaces and film growth temperatures (**Suppl. 2.5.1.**). The further decrease observed in the spin transition yield during the thermal cycling of the tetragonal film can be attributed to a partial conversion of the amorphous component into the tetragonal crystalline polymorph (see below in the description of the electrical sensing of the integrated devices).

Interestingly, for the as-sublimed films an instantaneous Soft X-ray Induced Excited Spin State Trapping (SOXIESST) effect is also observed at low temperatures, which is more sensitive for the as-sublimed tetragonal polymorph (below ~90 K), as compared with the triclinic one (below ~40 K). Moreover, the thermally cycled tetragonal film also exhibits a reverse-SOXIESST effect upon isothermal successive X-ray irradiation periods (**Suppl. 2.5.3.2.**). This is a very rare and unprecedented behavior for this kind of materials, which underlines that this polymorph is highly sensitive to light irradiation and suggests that the tetragonal polymorph is SCO active.

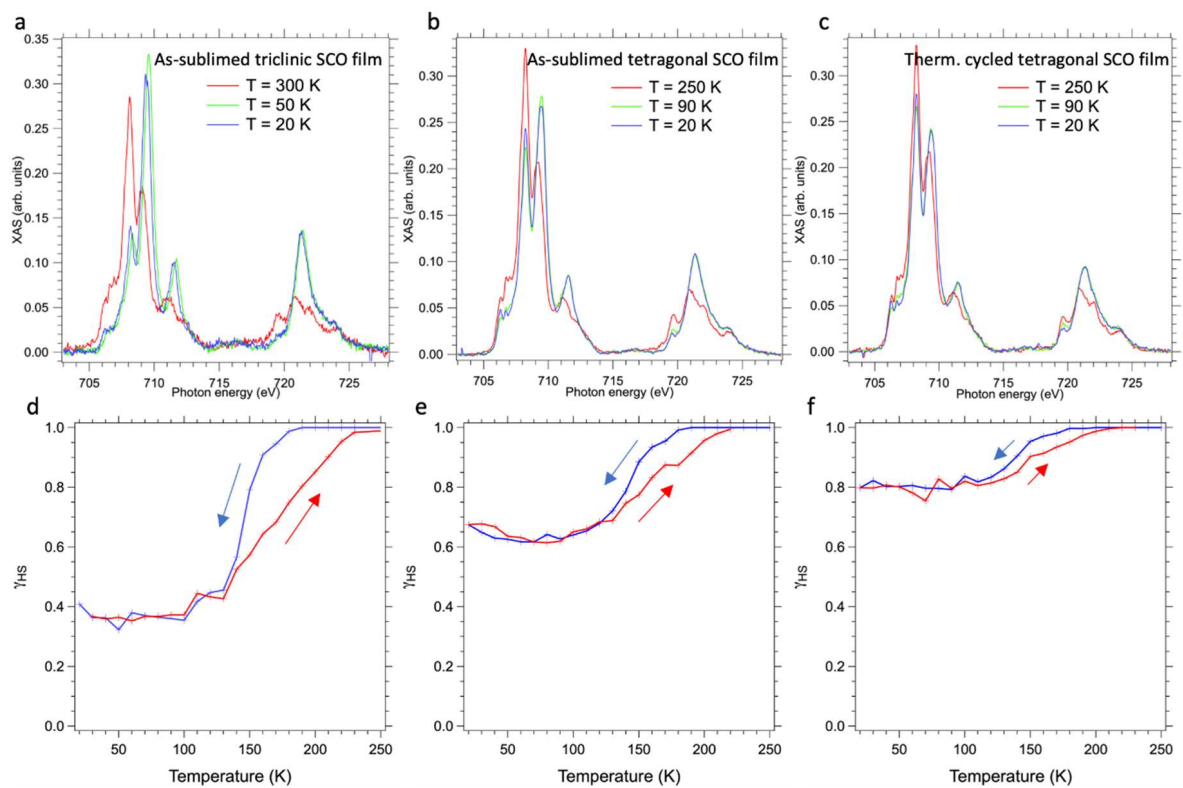

**Figure 2.** a) XAS spectra collected at 300 K, 50 K and 20 K in the Fe $L_{2,3}$ edges energy range for an as-sublimed [Fe(Pyrz)$_2$] 100 nm thick film deposited at 100 °C on SiO$_2$ and d) calculated HS fraction from XAS spectra collected at each temperature for both cooling (blue line) and heating (red line) processes in a temperature range varying from 250 K to 20 K in 10 K steps. b) XAS spectra collected at 250 K, 90 K and 20 K in the Fe $L_{2,3}$ edges energy range for an as-sublimed [Fe(Pyrz)$_2$] 75 nm thick film deposited at -90 °C on SiO$_2$ and e) calculated HS fraction from XAS spectra collected at each temperature for both cooling (blue line) and heating (red line) processes in a temperature range varying from 250 K to 20 K in 10 K steps. c) XAS spectra collected at 250 K, 90 K and 20 K in the Fe $L_{2,3}$ edges energy range for a thermally cycled [Fe(Pyrz)$_2$] 75 nm thick film deposited at -90 °C integrating a horizontal CVD-graphene device and f) calculated HS fraction from XAS spectra collected at each temperature for both cooling (blue line) and heating (red line) processes in a temperature range varying from 250 K to 20 K in 10 K steps.

*Electrical transport properties of the in-contact SCO/graphene devices. Thermal switching.* Temperature dependent electrical transport measurements were performed in DC mode with a four-probe configuration as schematized in **Figure 1a** (**Suppl 4.1.**). As reference device, pristine CVD-graphene was measured under the same conditions (**Suppl. 4.2.1.**). Regarding the triclinic polymorph of [Fe(Pyrz)$_2$], the measurements were performed on devices integrating thin films of this molecular material deposited onto the CVD-graphene at 100 °C. We noted that changes in the transport properties were detectable for devices with a SCO layer of the order of 100 nm thick. So, we focused our study on a device with a 130 nm thick SCO layer. In this up to 11 consecutive full thermal cycles were run (**Suppl. 4.1.1. and 4.3.1.**). We observed that the thermal dependence of the electrical resistance mimics the thermal hysteretic magnetic behavior reported for the pure triclinic polymorph (a sharp transition at ~145 K upon cooling ($T_{1/2}\downarrow$) and a more progressive transition at ~185 K upon heating ($T_{1/2}\uparrow$); **Figure 3a-c**).[9,28–30]. In addition, a second resistance transition at ~60 K emerged upon thermal cycling, producing a separation in the resistance values between cooling and heating processes in the temperature range 60 K - 100 K (**Suppl. 4.3.1.**). As described below, this feature can be assigned to the appearance of the tetragonal polymorph. In fact, this metastable phase arises from a structural conversion of the triclinic polymorph upon thermal cycling during the electrical measurements, as demonstrated by SXRD (green line in **Figure 1b,c**).

As far as the tetragonal polymorph is concerned, several devices integrating thin films with different thickness were produced (85, 130 and 200 nm). We chose a device integrating an 85 nm thick [Fe(Pyrz)$_2$] film deposited onto CVD-graphene at -90 °C, since for lower thickness the spin transition was not detected (**Figure 3d; Suppl. 4.3.2.**). Upon thermal cycling (up to 17 cycles, one month of aging from the 16 cycle) the device exhibits high robustness. Thus, excepting some small changes in the behavior during the first thermal cycles (1-4), the behavior does not vary upon further cycling. It is characterized by an abrupt increase in the resistance (~20 Ω) at ~60 K during the cooling process (**Figure 3e**) and a very progressive decrease during the heating process finishing at ~190 K. This behavior constituted the first evidence of the thermal SCO transition experienced by the [Fe(Pyrz)$_2$] tetragonal polymorph film. Note that the small changes observed in the first thermal cycles are due to the progressive conversion of the amorphous component, initially present in the as-sublimed film, into the crystalline tetragonal polymorph, as demonstrated by SXRD (black line in **Figure 1b,c**) and XAS measurements (**Figure 2c,e**; **Suppl. 2.5.2.**). This structural evolution upon cycling was also characterized through Atomic Force Microscopy (AFM), which showed an increase in the average crystal grain sizes (**Suppl. 2.3.4.**).

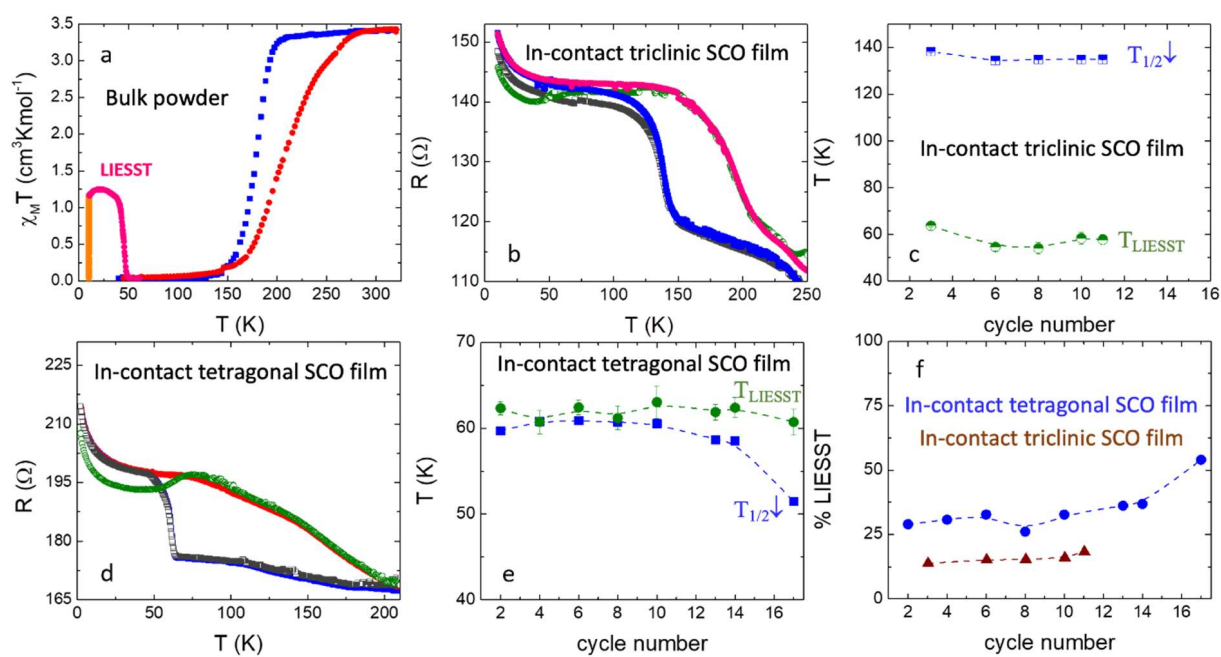

**Figure 3.** a) [Fe(Pyrz)$_2$] SCO molecule respective bulk powder first thermal cycle magnetic susceptibility measurement showing both thermal (blue line – cooling, red line – heating) and light induced SCO transitions (orange line – excitation, pink line – thermal relaxation). b) Temperature dependence of the resistance in an in-contact CVD-graphene horizontal device integrating a 130 nm thick film of [Fe(Pyrz)$_2$] in its triclinic polymorph in the range between 2 K and 250 K showing thermal cycles 2 (blue-filled squares line - cooling process and red-filled circles line - heating process) and 3 (empty grey squares line - cooling process and empty green circles line - heating process after light irradiation) respectively, and its c) calculated thermal

SCO transition temperature (half-filled blue squares line) during the cooling process ($T_{1/2}\downarrow$) and LIESST effect relaxation temperature (half-filled green circles line) during the heating process ($T_{LIESST}$) on cycles 3, 6, 8, 10 and 11 respectively. d) Temperature dependence of the resistance in an in-contact CVD-graphene horizontal device integrating an 85 nm thick film of [Fe(Pyrz)$_2$] in its tetragonal polymorph in the range between 2 K and 250 K showing thermal cycles 5 (blue-filled squares line - cooling process and red-filled circles line - heating process) and 6 (empty grey squares line - cooling process and empty green circles line - heating process after light irradiation) respectively, and e) its calculated thermal SCO transition temperature (blue-filled line) during the cooling process ($T_{1/2}\downarrow$) and LIESST effect relaxation temperature (green-filled circles line) during the heating process ($T_{LIESST}$) (on cycles 2, 4, 6, 8, 10, 13, 14 and 17 (after one month of aging). f) Calculated LIESST effect yield for different cycles in the two different in-contact devices prepared (brown-filled triangles line for b) device and blue-filled circles line for d) device).

To finish this section, we will show here that in these devices the pure tetragonal polymorph can quantitatively evolve to the triclinic one by annealing (**Suppl. 4.3.5.**). Thus, a 130 nm thick tetragonal polymorph film device was annealed at 90 °C for 7.5 h in total. In **Figure 4a** one can see a clear change in the transport properties when comparing a pre-annealing thermal cycle and a post-annealing one, which is consistent with a conversion between the two polymorphs ($T_{1/2}\downarrow$ of the tetragonal polymorph, appearing at ~60 K, disappears upon annealing at expenses of the appearance of the triclinic polymorphs at ~125 K). Notice that, in contrast to what is observed in the as-sublimed triclinic film, this annealing process results in a more progressive thermal SCO transition and a slight shifting of $T_{1/2}\downarrow$ and $T_{1/2}\uparrow$ to lower temperatures. By AFM one observes that this crystallographic change is accompanied by a drastic increase of both grain sizes and roughness both in a film annealed at different times (**Figure 4b**), and in the device (**Suppl. 2.3.4.**). Finally, one observes that by performing in this device various consecutive thermal cycles after the annealing, a small recovery of the tetragonal polymorph in the film is obtained (**Suppl. 4.3.5.**), as sensed for the as-sublimed triclinic phase device (**Suppl. 4.3.1.**). These observations are indicative of some reversibility in the conversion between the two polymorphs (partial from triclinic to tetragonal upon thermal cycling, and full from tetragonal to triclinic upon annealing).

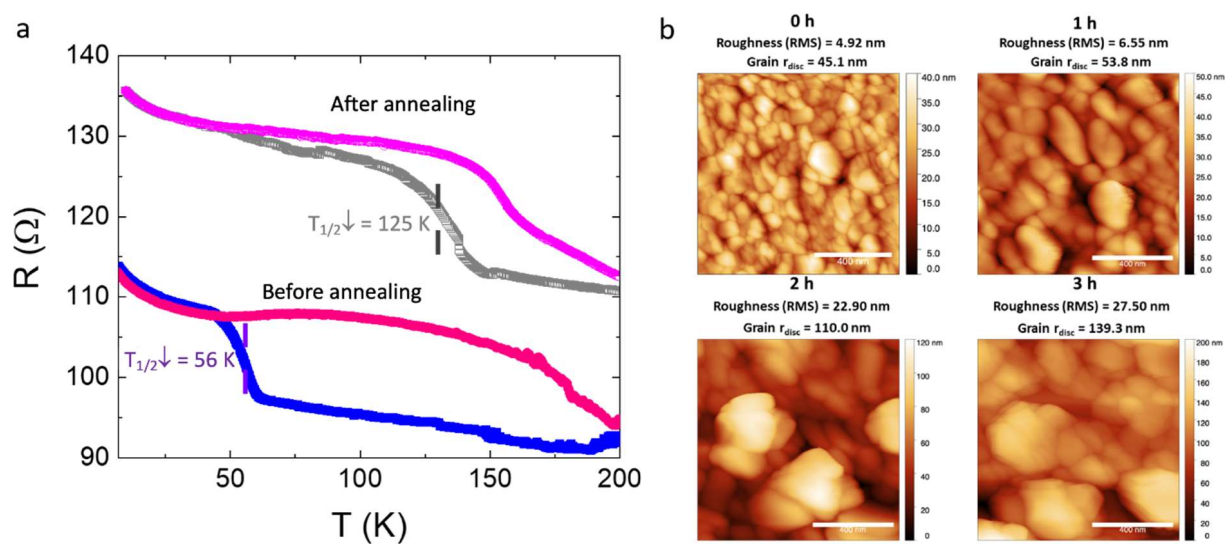

**Figure 4**. a) Temperature dependence of the resistance in the in-contact SCO/graphene device integrating a 130 nm thick film of [Fe(Pyrz)$_2$] in tetragonal polymorph for the third thermal cycle before (cycle 3, blue line – cooling process, red line – heating process) and the first thermal cycle after (cycle 7, grey line – cooling process, pink line – heating process) a total 7.5 h annealing at 90 ºC in the range between 2 K and 200 K. The main thermal SCO transitions sensed during the cooling process for each represented cycle are indicated as $T_{1/2}\downarrow$. b) AFM images collected of [Fe(Pyrz)$_2$] 40 nm thick films deposited on SiO$_2$/Si substrates at -90 °C and annealed in an Ar filled glovebox at 80 °C for (up-left) 0 h, (up-right) 1 h, (down-left) 2 h and (down-right) 3 h. The respective statistical value of the roughness in terms of RMS is indicated in the upper part of each AFM image.

*Electrical transport properties of the in-contact SCO/graphene devices. Photoinduced switching.* By irradiating the prepared devices at 2 K with a green laser ($\lambda$ = 532 µm, P = 28.2

µW·mm$^{-2}$), a decrease in the resistance was detected which was attributed to a partial LIESST effect (**Figure 3b,d,f**). By irradiating during 60 min the triclinic polymorph device, this change in resistance represents ca. 15% as compared with the change observed for the thermal one. Note that the only previous report in which a SCO/graphene device exhibits LIESST effect was based on films of this polymorph.[28] In that case higher light-induced spin conversion was achieved with a red laser (ca. 70% after 60 min), but at expenses of using a much higher power (ca. 20 times larger), thus limiting the device durability in terms of molecular film integrity.

By keeping the same experimental conditions for irradiating the tetragonal-based device, a spin conversion up to 50% in thermal cycles > 16 (one-month aging process) was accomplished after irradiating for 5 min (**Figure 3f**), tending to saturate for longer irradiation times. This indicates that the tetragonal polymorph is more susceptible to respond to light irradiation. In fact, this device shows an unprecedently fast response to LIESST effect as compared with the triclinic polymorph one, becoming highly useful for light addressable electronic or spintronic devices. In view of this key improvement, from now on we will focus on switching devices based on this polymorph.

*Electrical transport properties of the in-contact SCO/graphene devices. Influence of the gate voltage.* To analyze the effect of an applied gate voltage on the device based on the tetragonal polymorph two additional consecutive thermal cycles were performed at positive (70 V) and negative (-70 V) voltages (**Suppl. 4.3.3.**). In this experiment no variation on T$_{1/2}$↓ was found, independently of the gate voltage applied. However, a drastic decrease of the resistance change (~75 %) was observed when these gate voltages were applied. This decrease arose from a shift in the Fermi level, which separates it from the Dirac point.[32] Furthermore, the change experienced in the device sensitivity to the spin transition presented a similar value for both gate positive and negative voltages. Thus, we can conclude that the Fermi level is close to the Dirac point, exhibiting the device a proper sensitivity to the spin transition, without applying any electric field.

*Preparation and study of contactless switching devices based on the SCO tetragonal polymorph.* With the aim of checking if the spin transition can still be sensed when the SCO system is not in direct contact to CVD-graphene, we fabricate an alternative device configuration based on the deposition of an inert PMMA layer in between both active elements of the device (**Figure 5a**). This approach was already used for sensing the thermal spin transition in a bulk single crystal of a different SCO compound.[18] In our case, the device contains a 30 nm thick layer of PMMA resist between the CVD-graphene and a 100 nm thick [Fe(Pyrz)$_2$] tetragonal polymorph sublimed film (**Suppl. 3.3.2.**). The electrical transport measurements indicate that both the thermal SCO transition and LIESST effect can be seen in this contactless device (**Figure 5b,c,e**; **Suppl. 4.3.4.**). This proves the successful long-range sensing of the intrinsic SCO properties of the molecular film in the tetragonal polymorph. In particular, the change in the resistance accompanying the spin transition decreases from ~20 Ω in the above-described in-contact device to ~4 Ω in the present contactless one. This is not an unexpected outcome since the interactions between the two active components have to be reduced upon the insertion of the separating PMMA layer. More interesting is the substantial enhancement of the LIESST effect (**Figure 5b,c,e**). Thus, using the same conditions as for the in-contact devices (irradiation during 5 min at 2 K with a green laser: $\lambda$ = 532 μm, P = 28.2 μW·mm$^{-2}$), a complete photoinduced spin transition (close to a 100 % yield in thermal cycles > 11) was observed (**Figure 5d**). This outstanding yield may be related with the elastic properties of the PMMA resist,[33] since thanks to its flexibility the volumetric change experienced by the SCO compound during the light-induced spin transition could be facilitated. In fact, the yield of the LIESST effect drastically decreased to ~35 % (cycle 15 in **Figure 5c**, **Suppl. 4.3.4.**) when submitting the device to a two-month aging process (at room temperature and inert atmosphere). Under these conditions, the mechanical properties of PMMA do change,[34–37] experiencing a hardening and decrease of volume, that may difficult the volumetric change associated to the spin transition. Intriguingly, this aging process seems to enhance the sensing of the thermal spin transition (the resistance change in the graphene increases from ~4 Ω to ~8 Ω, see **Figure S43**), an effect that could be related with the better transmission of the strain generated in the SCO layer towards the graphene via the more rigid and thinner PMMA bridging layer.

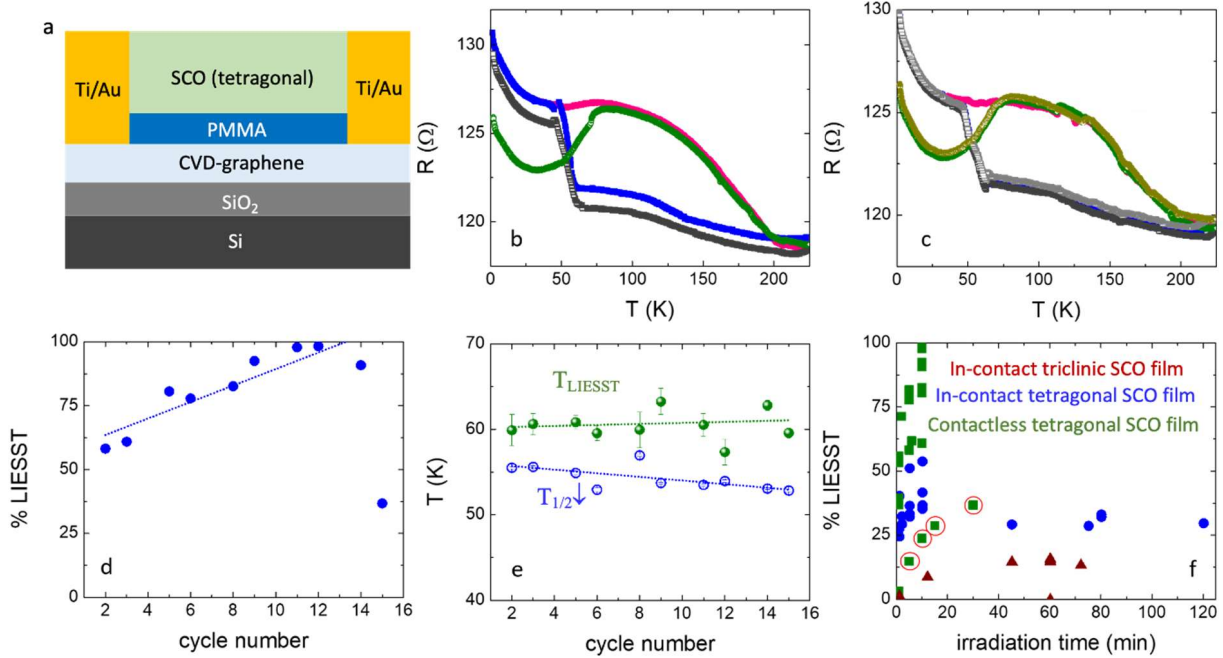

**Figure 5**. a) Vertical profile of the fabricated contactless SCO/graphene device integrating a [Fe(Pyrz)$_2$] film for the temperature dependence of the electrical transport measurements and a PMMA interlayer. Temperature dependence of the resistance in a contactless CVD-graphene horizontal device integrating a 100 nm thick film of [Fe(Pyrz)$_2$] in its tetragonal polymorph deposited on top of a 30 nm thick PMMA interlayer separating it from the graphene in the range between 2 K and 225 K showing b) thermal cycles 4 (blue-filled squares line - cooling process and red-filled circles line - heating process) and 5 (empty grey squares line - cooling process and empty green circles line - heating process after light irradiation) respectively, and c) thermal cycles 9 (empty grey squares line - cooling process and empty green circles line - heating process after light irradiation), 10 (blue-filled squares line - cooling process and red-filled circles line - heating process) and 11 (half-filled grey squares line - cooling process and half-filled gold circles line - heating process after light irradiation) respectively. d) Calculated LIESST effect yield of the contactless device in light irradiated cycles 2, 3, 5, 6, 8, 9, 11, 12,

14 and 15. e) Calculated thermal SCO transition temperature (empty blue squares line) during the cooling process ($T_{1/2}\downarrow$) and LIESST effect relaxation temperature (green-filled circles line) during the heating process ($T_{LIESST}$) on cycles 2, 3, 5, 6, 8, 9, 11, 12, 14 and 15. In a) and b) the cycles where the devices have been green-light irradiated while held at 2 K are indicated with a green ray picture. f) Comparison between the LIESST effect yield observed for each prepared type of device as a function of irradiation time including the data obtained for different thermal cycles in each case. In f) the red-circled data of the contactless tetragonal SCO film device correspond to the thermal cycle measured after a two-month aging.

## 3. Conclusion

In this work, robust SCO/graphene devices electrically sensing both thermal and photo-induced spin switching have been fabricated, using for the first time a contactless configuration. A typical device of this kind is formed by a 30 nm thick layer of PMMA resist between the CVD-graphene and a 100 nm thick [Fe(Pyrz)$_2$] film of the tetragonal polymorph. These hybrid devices show unprecedented features including a quantitative and fast (just a few minutes) light-induced spin transition at low temperature (LIESST effect response). To reach this final outcome a novel sublimation protocol has been developed to directly integrate thin films of [Fe(Pyrz)$_2$] molecule on horizontal graphene devices. This method has involved in a first step to control the growth of the two known crystalline polymorphs of this molecule (triclinic and tetragonal) at will. This has enabled us to prepare devices integrating either as-sublimed triclinic or tetragonal polymorphs over graphene. In a second step, these devices have allowed us to sense the different spin transition properties in the two polymorphs by measuring the electrical transport properties of the graphene. Using this approach both thermal and light-induced spin transitions have been probed for the first time for the tetragonal polymorph. Interestingly, this behavior has shown better performances than the reported ones for the triclinic polymorph (both in terms of thermal hysteresis and LIESST yield). In fact, before the present study the SCO transition in the tetragonal "metastable" polymorph was considered to be inactive and all the previous studies were focused on the triclinic "stable" polymorph.[9,23,28–30,38–40] That was mainly due to the high sensitivity of the tetragonal polymorph to light irradiation, alongside its very low spin transition temperature and its intrinsic structural instability above room temperature (undergoing a conversion towards the triclinic polymorph). Here, this elusive polymorph has been properly isolated demonstrating a fast light-induced spin switching with more robustness and efficiency when integrated in the in-contact devices as compared with the triclinic-based ones (50% compared to 15%, after irradiating for 5 and 60 min, respectively). In a final step, a further improvement of the photo-induced spin switching has been achieved (ca. 100% after 5 min) by inserting a flexible PMMA layer, which should facilitate the volume change of the crystalline SCO film upon the spin transition in the SCO/graphene heterostructure. This methodology is fully compatible with the microelectronics industry, thus opening a wide range of possibilities for further integration of these robust and sublimable SCO molecules in thermal or light-addressable spintronic and straintronic devices based on 2D materials. We can imagine for example to spatially-control the strain over the 2D material with high-resolution (<20 nm) via an active local manipulation of the spin state in the SCO layer, both at low (induced by light) or intermediate/high temperatures (induced by thermal heating). This thermal or light-induced strain generated by the SCO molecules is complementary to the electric field-induced strain generated by applying a voltage between a gate electrode and a suspended 2D membrane,[41] and should provide examples of strain-engineered 2D materials of interest to transport information via magnons, excitons, and phonons.

## 4. Experimental Section/Methods

*Bulk powder synthesis and SQUID characterization:* The bulk material of this molecule was synthesized following the procedure previously described in the literature obtaining a crystalline white powder.[7] Magnetic susceptibility measurements were performed via

Superconducting Quantum Interference Device (SQUID magnetometer, model Quantum Design MPMS-XL-7) technique in direct current (dc) mode, 1000 Oe magnetic field and cooling/heating rates of 1 K·min$^{-1}$ in a range between 300 K and 40 K. LIESST effect characterization was done irradiating the powder at 5 K using a 532 nm LASER with a power of 11.2 mW followed by a heating ramp of 0.3 K·min$^{-1}$ up to 60 K.

*Thin film Sublimation:* Molecular films of [Fe(Pyrz)$_2$] with thicknesses ranging from 10 nm to 250 nm were produced by sublimation in HV of the readily synthesized bulk material using a CREATEC Molecular Beam Epitaxy (MBE) customized system in a Clean Room Class 10000. The available thermal heater (Knudsen Cell) inside the HV growth chamber of the equipment was filled with 500 mg of the white crystalline powder in a fitting quartz crucible and heated at 145 °C in a 5·10$^{-8}$ mbar pressure HV to reach an average deposition rate of 0.4 Å·s$^{-1}$ on the substrates sensed with a Quartz Crystal Microbalance present at the same distance as the substrate from the molecular source (approximately 50 cm). To selectively control crystallographically the molecular films the deposition temperature was held either at -90 °C or 100 °C respectively by introducing liquid nitrogen into the sample plate holder cavity or using the thermal heater of the sample plate holder.

*IR spectroscopy:* The chemical integrity of the prepared molecular films of [Fe(Pyrz)$_2$] was confirmed via IR spectroscopy (Fourier Transformation-Infrared Spectrometer NICOLET 5700 from Thermo Electron Corporation) comparing with the bulk material. For the bulk powder, a module was used that allowed the detection of the transmitted light after passing through [Fe(Pyrz)$_2$] bulk material embedded in a KBr pellet. For the thin film, a second module to measure the transmittance of the reflected light from 100 nm thin films deposited in each crystallographic phase on a 3 cm x 3 cm Au coated glass substrate. All spectra, collected between 650 cm$^{-1}$ and 4000 cm$^{-1}$, were normalized to facilitate their comparison.

*Raman spectroscopy*: Composition of the molecular films was analyzed through Raman spectroscopy using a HORIBA LabRAM HR Evolution equipped with a 473 nm Laser beam with a power of 1.08 mW·µm$^{-2}$. The bulk material was placed directly on a glass slide and measured whereas 100 nm thick molecular layers were deposited at -90 °C and 100 °C respectively on SiO$_2$/Si substrates for the thin film characterization. All spectra, collected between 1000 cm$^{-1}$ and 3200 cm$^{-1}$, were normalized to facilitate their comparison.

*AFM imaging:* The topography of all the thin films of [Fe(Pyrz)$_2$] characterized in this work were performed using a Bruker (former Veeco) Nanoscope IVa Multimode Scanning Probe Microscope in tapping mode. The images were collected in sizes of either 1 µm x 1 µm or 5 µm x 5 µm and treated using the Gwyddion program. Roughness values were acquired statistically from the Root Mean Square (RMS) value calculated by this program. The average grain radii (Grain r$_{disc}$) were calculated from the average grain area given by the program selecting a certain thickness threshold or from manually selected crystal grains.

*Powder X-Ray Diffraction (PXRD)/SXRD:* The SXRD pattern of [Fe(Pyrz)$_2$] films integrated in the CVD graphene-based horizontal devices different deposition temperatures before and after transport properties characterization was collected at room temperature using a PANalytical Empyrean X-ray platform with PIXcel detector equipped with a Cu Kα X-ray source. PXRD pattern of the bulk powder introduced inside a glass capillary was also collected using the same equipment for comparison. All patterns, collected between 5 ° and 35 ° in 2θ, were normalized to facilitate their comparison.

*XAS:* XAS characterization of thin films as-sublimed on Si/SiO$_2$ and after the performance of various consecutive thermal cycles in a device was performed at Boreas beamline in ALBA synchrotron. The experiment consisted in the collection of multiple energy scans focused at the Fe $L_{2,3}$ edge region where a set of temperatures were applied onto each sample using total electron yield (TEY) mode and the lowest possible photon flux (intensity = 0.040 nA) enabling the obtention of a good signal to noise ratio. For the LIESST effect characterization, the sample temperature was set to 2 K and then irradiated with a green Laser (532 nm, 35 mW) for 60 min in 10 min steps. XAS spectra were collected between each of the steps only exposing the sample to X-rays during these periods to avoid as much as possible the SOXIESST effect. All collected spectra for this study were processed for their analysis through background subtraction and normalization. The HS fraction for all cases was calculated from a linear combination of the normalized spectra of the bulk material for pure LS and the one for the pure HS.

*Devices transport properties characterization:* The measurement of the graphene electrical transport properties in each device manually wired bonded to a chip carrier fitting into the cane of our Physical Properties Measurement System (PPMS) model Quantum Design PPMS-9 was performed in DC mode with a four-probe configuration. At this respect, two parallel pairs of electrodes with the same distance between them (5 mm or 7 mm) were used at the same time; supplying DC intensity current through one pair of them whereas the other one sensed the voltage drop. Using this configuration different studies on the variation of the transport properties of the devices as a function of temperature were considered. Mainly, these measurements consisted in studying this change in the electrical response of the devices with and without laser irradiation when reaching 2 K during the cooling process. For the thermal dependence study, samples were cycled in a range between 250 K and 2 K (cooling/heating rates of 1 K·min$^{-1}$) for at least 10 times to analyze both cyclability and reproducibility of the electrical readout of the SCO characteristics. For the LIESST effect sensing on the devices, the samples were held at 2 K after reaching this temperature during the cooling process in some randomly selected thermal cycles and irradiated with a 532 nm laser with a power of 28.2

µW·mm$^{-2}$ for at least 5 min via an optical fiber going across the PPMS cane directly pointing to the [Fe(Pyrz)$_2$] film deposited as top layer of the device. Once the irradiation period ended, the laser was switched off and the heating process restarted in the conditions specified above. For the gate-voltage study, three consecutive full thermal cycles in the same conditions described previously for the electrical transport measurements as a function temperature were performed in a device connected to the PPMS system to allow its electrical gating and applying respectively 0 V, 70 V and -70 V voltages in each of the three consecutive thermal cycles. Both different devices fabrication and transport properties measurements are further and extensively described in Suppl. 3. and 4.

**Supporting Information**
Supporting Information is available from …. or from the author.

**Acknowledgements**

We acknowledge the financial support from the European Union (ERC AdG Mol-2D 788222, Research Executive Agency (REA)-FETOPEN COSMICS 766726), the Spanish MICINN (2D-HETEROS PID2020-117152RB-100, SUPERSCO PID2020-117264GB-100, co-financed by FEDER, Grant PID2020-117264GB-I00 funded by MCIN/AEI/10.13039/501100011033 and Excellence Unit "María de Maeztu", CEX2019-000919-M), and the Generalitat Valenciana (Prometeo program and PO FEDER Program, ref IDIFEDER/2018/061 and IDIFEDER/2020/063). M.G.-E. acknowledges the support of a fellowship FPU15/01474 from MINECO. R.C. acknowledges the support of a fellowship from "la Caixa" Foundation (ID 100010434), LCF/BQ/PR19/11700008 and the Generalitat Valenciana (SEJIGENT/2021/012T). F. J. V. M. acknowledges the support of the Generalitat Valenciana (APOSTD/2021/359). J. A. R. acknowledges the support of a grant PID2019-106147GB-I00 funded by MCIN/AEI/10.13039/501100011033. All XAS experiments were performed at Boreas beamline at ALBA Synchrotron with J.H.-M. in both proposal and in-house experiments.

The authors thank Yahya Shubbak for his support in the fabrication of some of the thin films used in the XAS characterization and Ángel López-Muñoz for his technical support.